\documentclass[a4paper,10pt]{scrartcl}
\usepackage[nolist]{acronym}
\usepackage{amsmath}
\usepackage{authblk}
\usepackage{graphicx}
\usepackage{hyperref}
\usepackage[utf8]{inputenc}
\usepackage{natbib}

\addtokomafont{disposition}{\rmfamily}

\setcapindent{0pt}

\begin{acronym}
  \acro{APE}{absolute percentage error}
  \acro{EDA}{electronics design automation}
  \acro{IQR}{interquartile range}
  \acro{IC}{integrated circuit}
  \acro{PPA}{power, performance, and area}
  \acro{ROM}{read-only memory}
  \acro{RF}{register file}
  \acro{SoC}{system-on-chip}
  \acro{SRAM}{static random access memory}
  \acro{VT}{voltage threshold}
  \acro{SRAM}{static random access memory}
  \acrodefplural{SRAM}[SRAMs]{static random access memories}
  \acrodefplural{SoC}[SoCs]{systems-on-chip}
\end{acronym}

\begin{document}
\title{Predicting Memory Compiler Performance Outputs using Feed-Forward Neural Networks}
\date{}

\author[1,2]{Felix Last}
\author[2]{Max Haeberlein}
\author[1]{Ulf Schlichtmann}
\affil[1]{Department of Electrical and Computer Engineering, Technical University of Munich}
\affil[2]{Intel Deutschland GmbH, Neubiberg, Germany}

\maketitle

\begin{abstract}
  Typical semiconductor chips include thousands of mostly small memories. As memories contribute an estimated 25\% to 40\% to the overall \ac{PPA} of a product, memories must be designed carefully to meet the system's requirements. Memory arrays are highly uniform and can be described by approximately 10 parameters depending mostly on the complexity of the periphery. Thus, to improve \ac{PPA} utilization, memories are typically generated by memory compilers. A key task in the design flow of a chip is to find optimal memory compiler parametrizations which on the one hand fulfill system requirements while on the other hand optimize \ac{PPA}. Although most compiler vendors also provide optimizers for this task, these are often slow or inaccurate. To enable efficient optimization in spite of long compiler run times, we propose training fully connected feed-forward neural networks to predict \ac{PPA} outputs given a memory compiler parametrization. Using an exhaustive search-based optimizer framework which obtains neural network predictions, \ac{PPA}-optimal parametrizations are found within seconds after chip designers have specified their requirements. Average model prediction errors of less than 3\%, a decision reliability of over 99\% and productive usage of the optimizer for successful, large volume chip design projects illustrate the effectiveness of the approach.
\end{abstract}

\newpage
\section{Introduction}

\subsection{Motivation}
\label{sec:motivation}
For the past five decades, the electronic chip industry persistently fulfilled on the exponential frequency and power improvements initially forecasted by Gordon Moore in the 1960s. However, maintaining steady improvements at such ever-increasing pace becomes more and more challenging on today's advanced technology nodes, where new physical phenomena and limitations in lithography resolution prohibit mere continuation of the current practice. Instead, \acf{PPA} trend continuation at aggressively scaled sub-20-nm nodes requires substantial innovation of \ac{EDA} tools.

Owing to their sheer count on today's chips, memories such as \acp{SRAM}, \acp{ROM}, and \acp{RF} have significant impact on the \ac{PPA} on modern \acp{IC}. Because different requirements apply to each memory, such as bit size and the number of I/O ports, there is no single blanket solution for optimal memory design. Instead, each memory instance must be fine-tuned separately in order to achieve overall product \ac{PPA} targets. Aggravatingly, focus often shifts during the design process, with initial priority given to minimal area until power concerns begin to dominate later design cycles. Consequently, techniques are needed to select optimal memory designs for given requirements to serve the short turnaround times and flexibility requirements of an increasingly dynamic design environment.

Memory compilers are the most common choice for generating memories in modern \ac{IC} design flows \citep{Guthaus.2016}. These tools, which are typically provided by the manufacturing foundry or external vendors, provide parametrizable libraries of memories which are verified and characterized for the respective technology node. Given parameters such as the number of words, the word width, as well as a multitude of architectural parameters, a memory compiler returns the netlists, RTL codes, and other artifacts required both for the \ac{IC} design flow and eventually for manufacturing. Moreover, the compiler produces accurate \ac{PPA} estimates for the memories. However, with a growing number of architectural parameters which target the improvement of various \ac{PPA} aspects, complexity of memory compilers also increases significantly.

The vast number of options chip designers have to configure makes it difficult to determine those settings most suitable for the overall product's \ac{PPA} targets. This is especially true since most of the compiler input parameters are architectural parameters, for which values can not be derived directly from system requirements, yet have significant effects on \ac{PPA}.

Interactions with system parameters make the task of selecting architectural parameters significantly more complex. To name one example, a large macro may have to be split into multiple banks along the wordline in order to achieve frequency requirements, whereas unnecessary banking beyond matching frequency should be avoided to keep area minimal. Therefore, the number of banks - an architectural parameter - must be carefully tuned for each individual memory in accordance with its size and target frequency.

Compiler run times of more than 30 minutes prohibit manual or automated trial-and-error search of optimal compiler input parameters. Moreover, for a given set of system parameters, multiple compilers may be available on a single technology node, further and significantly expanding the search space. Lastly, during architecture exploration phase, where different technology nodes are compared, achievable \ac{PPA} must be estimated for all applicable compilers even across vendors and process technologies.

In practice today, no standard solution to optimizing memory compiler input parameters has been established. Instead, parameter selection is often done by experts with significant experience of working with the respective compilers. However, expert knowledge is challenged by rapid compiler update cycles and a wide range of available technology nodes, limiting the applicability of previously learnt heuristics.

Chip designers and compiler experts also frequently rely on optimization tools shipped by compiler vendors. Aside from the clear limitation of these tools' applicability to the respective technology node, which obstructs comparison across nodes and vendors, their accuracy and run times are often not satisfactory, a result of over-simplified models or dependence on exhaustive compiler executions. \ac{IC} design teams are further forced to rely on the vendor to provide an optimizer and to keep it up-to-date, while they are granted only limited insight into the reliability of its results.

\subsection{Related work}
\label{sec:related-work}
While, to the best of our knowledge, no published literature exists on finding memory compiler input parameter which optimize \ac{PPA}, several optimization problems in the \ac{EDA} space exhibit similar characteristics. The generalized goal of such optimization problems is to find optimal parameters of electrical components with regard to some target dimension such as \ac{PPA} or yield. The limiting factor of such optimization is usually a costly evaluation of solutions, which in our case involves memory compiler execution and otherwise mostly requires expensive simulations.

\cite{Wang.2018} distinguish three approaches for yield optimization: Monte Carlo-based, corner-based and model-based optimization. As Monte Carlo-based techniques require an extensive amount of solution evaluations, most published work focusses on extensions which reduce the total number of required evaluations or simulations. Corner-based approaches, on the other hand, focus on finding parameters which optimize worst-case performance, which requires significantly fewer evaluations, but may lead to over-design and consequently unused potential with regard to the optimization target. Lastly, model-based approaches substitute the simulator (or, for our task, the memory compiler) during optimization and enable cheap evaluation of solutions.

\cite{Yao.2015} maximize \ac{SRAM} yield using a genetic algorithm which evaluates solutions using either a circuit simulation or a surrogate model trained on-the-fly, depending on an importance heuristic computed for each solution. Similarly, \cite{Wang.2018} also combine a model-based and a Monte Carlo-based approach for \ac{SRAM} yield optimization, exploiting Bayesian uncertainty estimates to simulate only those solutions where both uncertainty and estimated yield are high, while using an incrementally fitted model for other solutions.

Memory compiler vendors often provide optimizers to find optimal architectural parameters for given system parameters. A major shortcoming of these solutions is lack of transparency as neither accuracy nor methodology are published or shared with customers. The issue is magnified by bad accuracy owing to oversimplifications in model-based approaches such as linear models. Other solutions typically rely on linear interpolation of the \ac{PPA} of similar parametrizations in a large database which is expensive to construct and maintain. Such methods are more closely related to Monte Carlo-based methods. Finally, the use of vendor optimization tools requires system architects who wish to compare technology nodes to switch between several user interfaces which may provide \ac{PPA} outputs in different units or display different dimensions altogether.

Our proposed approach falls in the model-based category, which provides the advantage of fast solution evaluation without expensive simulations or compiler runs. Model-based approaches therefore support many or even exhaustive solution evaluations during an optimization run. However, according to \cite{Wang.2018}, model-based approaches involve a trade-off in terms of accuracy while requiring a large number of a-priori simulations to generate training data for model fitting. We argue that the latter drawback is acceptable when trained models are used not for a single optimization task, but for multiple designs and in multiple stages of the design process. Moreover, sufficient accuracy can be guaranteed by thorough and unbiased model evaluation. Lastly, as explained in detail in Section \ref{sec:decision-quality}, model accuracy does not need to be perfect as long as solutions selected based on model estimates are optimal. A single, final simulation (or compiler run) of the selected solution can ensure a reliable estimate of the target dimension (e.g. \ac{PPA} or yield).

\subsection{Proposed Solution Outline}
In order to tackle the challenge of finding memory compiler input parameters which optimize \ac{PPA} for a given memory, we propose fitting behavioral models of memory compilers. Realized as feed-forward neural networks, such models are capable of accurately predicting \ac{PPA} of a given compiler parametrization in extremely short time, thereby enabling efficient search of compiler options for given design requirements. This approach not only automates the memory selection and parametrization process, significantly reducing its complexity, but also improves \ac{PPA} results by enabling a much broader set of possible parametrizations to be evaluated. Applicable independently of vendor, the approach further facilitates rapid \ac{PPA} estimation and comparison of compiler vendors and technology nodes.

\subsection{Structure}
This work is structured as follows: In Section \ref{sec:proposed-solution} we detail the proposed solution. Herein, the optimizer framework is discussed separately from the behavioral models, which are described in terms of data structure, neural network architecture, and training procedure. We proceed to evaluate our approach in Section \ref{sec:evaluation}, dividing the analysis into model evaluation on the one hand and resulting decision quality assessment on the other. Finally, we summarize our findings and provide an outlook on future work in Section \ref{sec:conclusion}.

\section{Proposed Solution}
\label{sec:proposed-solution}
\begin{figure}[!htb]
  \centering
  \includegraphics[width=1\textwidth]{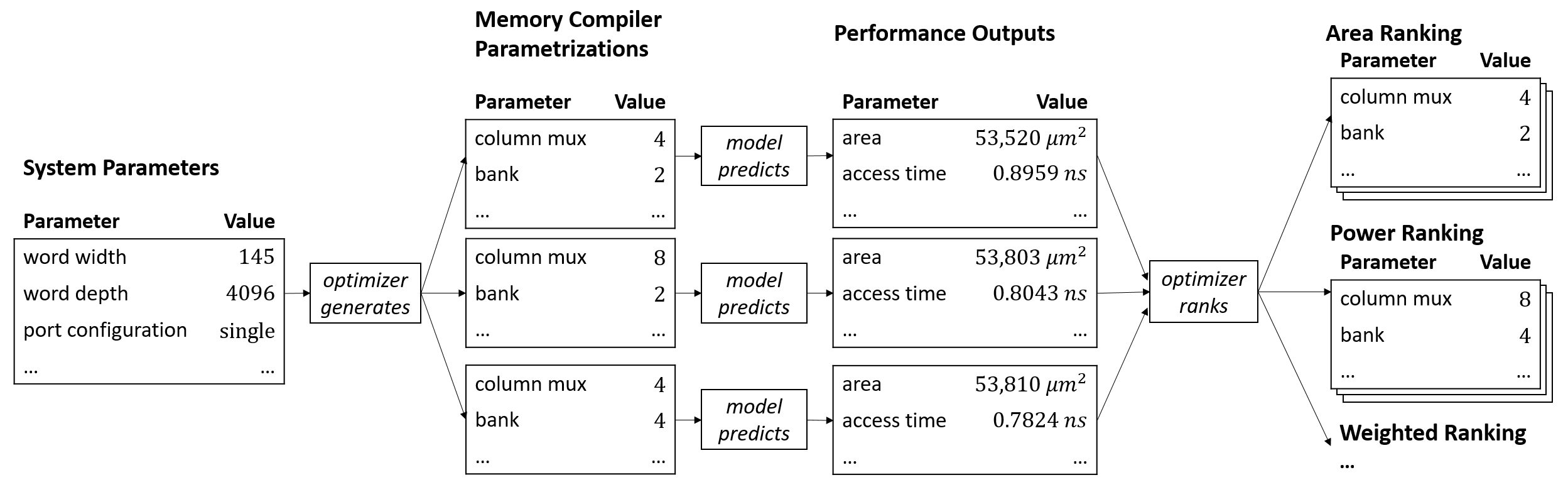}
  \caption{Schematic representation of the proposed solution architecture. Given system parameters, the optimizer generates all possible compiler parameter combinations. The behavioral models are used to predict the \ac{PPA} of each solution, before the optimizer removes solutions with insufficient frequency and ranks remaining solutions according to various \ac{PPA} criteria.}
  \label{fig:solution-overview}
\end{figure}

To the challenging task of finding optimal memory compiler input parameters for given system parameters, we propose a two-part solution: an optimizer framework which searches the parameter space, and neural networks trained as behavioral models of the memory compilers, which predict the \ac{PPA} of potential solutions. The overall architecture is presented schematically in Figure \ref{fig:solution-overview}.

\subsection{Optimizer}
\label{sec:optimizer}
To better understand the role of the behavioral models, we firstly introduce the memory optimizer, the interface through which chip designers access the behavioral models' predictions.

The optimizer aims to find a set of compiler input parameters which optimize \ac{PPA}. An optimization run is characterized primarily by the set of fixed and free compiler input parameters. Fixed parameters are predetermined by the chip designer in accordance with system parameters, the latter of which can be viewed as external requirements to the memory. In contrast, free parameters are subject to optimization and consist of mostly of architectural parameters, which control the internal architecture of the memory and significantly affect its \ac{PPA}.

The fixed parameters port configuration, word width, and word depth must always be specified. The port configuration, which determines the number of read and write ports of a memory, qualifies the set of compilers available for optimization. Similarly, word depth and word width may limit the set of available compilers because compilers have a fixed size range. Fixed parameters usually comprise other predetermined system parameters. For example, a memory with a separate voltage for the periphery is only viable when the \ac{IC} can provide separate power supplies, therefore chip designers typically fix the compiler's ``dual rail'' parameter in advance. However, system parameters may also remain undetermined, for example when comparing technology nodes during architecture exploration phase. When left undetermined, system parameters are added to the set of free parameters.

Architectural parameters, on the other hand, always remain free and subject to optimization. Each free parameter can assume one of multiple discrete values. The set of possible values is determined by the choice of compiler, as well as the specified word depth and word width. For example, a compiler may allow only up to two banks for a memory with small word depth, while requiring at least four banks for deeper memories.

The set of all possible parametrizations of a compiler is approximately in the order of magnitude of $10^7$. Clearly, it is not viable to execute memory compilers prophylactically to determine the \ac{PPA} of the full compiler parametrization space. The search space of a single optimization run, on the other hand, consists of all legal compiler parametrizations or solutions which match the set of fixed parameters, most importantly word depth and word width. The fixed parameters therefore constrain the search space to a hyperplane of the total parametrization space.

Another perspective is that the number and range of free parameters define the number of possible solutions. Each free parameter adds a dimension to the combinatorial search space of the respective compiler. For example, the search space of a compiler with four free parameters where each takes on one of three possible values consists of $3^4 = 81$ solutions. Given four available compilers, the total number of solutions is then $4 \times 81 = 324$. More free parameters and more available combinations can easily increase this number to multiple thousands.

Even for a single memory optimization task with fixed word depth and word width, long compiler run times of at least 30 minutes make exhaustive execution infeasible. Assuming a compiler run time of 30 minutes and parallel execution of 20 compiler processes, the time to evaluate $n$ possible solutions is given by $\frac{n}{40}$ hours. For the above example of 324 possible parametrizations, an exhaustive search directly using memory compilers would take over 8 hours while the evaluation of 1000 solutions would take 25 hours. Even in the unrealistic case of $n$ parallel compiler executions, exhaustive evaluation time is lower bounded by the compiler run time. To make matters worse, these numbers concern the optimization of only a single memory, one of possibly thousands of memories in an \ac{IC}.

Numerical optimization approaches, both statistical and deterministic, are based on function evaluation in a loop. However, any such method's convergence time is lower bounded by the function evaluation time. Besides designing optimization methods that minimize the required number of function evaluations, methods have been proposed to replace the direct, usually extremely expensive function evaluation by models. By using neural networks (or another behavioral model with low inference time) to predict the \ac{PPA} outputs of memory compilers, the evaluation of a possible solution becomes significantly faster and computationally cheaper. In fact, neural network inference is so fast that, in practice, exhaustive evaluation of the search space can be done in less than 5 seconds for most cases as more than 150 solutions can be evaluated per second.

For a better understanding of the optimization context, it is important to consider that compiler \ac{PPA} outputs are the result of characterization of the memory array for different process variations (e.g. typical, fast, slow, referred to as process corner) and under different operating conditions (voltage and temperature) \cite{Weste.2015}. A combination of process corner and environmental conditions is referred to as design corner. As such, five of six \ac{PPA} dimensions, precisely dynamic read power, dynamic write power, leakage, access time, and cycle time, but not area, are represented not as a single value per memory, but as a set of design corner-specific \ac{PPA} variables.

The optimizer results display as pairs of compiler parametrizations and respective \ac{PPA} outputs as predicted by the behavioral model. These pairs are organized in four separate lists, each ranked according to a different \ac{PPA} dimension. The ranking criteria of the lists are dynamic power, leakage, area, and a weighted sum of the former. All evaluated pairs appear in all four lists, albeit in potentially different ranking positions. The frequency dimension is not used for result ranking; instead, it is used as a threshold criterion to remove solutions which do not achieve product frequency requirements. A frequency exceeding requirements does not impact ranking of results in any way. As all \ac{PPA} dimensions except area consist of multiple \ac{PPA} variables (one for each design corner), a single \ac{PPA} variable must be determined per \ac{PPA} dimension which is used for ranking (or, in the case of cycle time, filtering) of results. The chip designer must therefore select one design corner per \ac{PPA} dimension to determine the relevant \ac{PPA} variable.

In addition to the three rankings based directly on \ac{PPA} dimensions, the weighted sum ranking linearly combines dynamic power, leakage and area dimensions of each result according to weighting factors defined on \ac{IC} product level. The specific \ac{PPA} variables used for this computation depend on the design corners chosen by the chip designer. Because the variables combined in this weighted ranking are on different scales and display different variances, a linear combination in original scale is not practical. For example, an area in the magnitude of thousands of square micrometers would otherwise have a much larger effect on the ranking than dynamic power, which varies on a scale which is around three orders of magnitude smaller. Therefore, before the weighted ranking can be computed, each variable must be brought to a common, comparable scale by means of standardization, i.e., subtracting the mean and dividing by standard deviation. Mean and variance are estimated based on known \ac{PPA} data as part of data preparation, which is discussed in Section \ref{sec:behavioral-model}.

\subsection{Behavioral Model}
\label{sec:behavioral-model}
As discussed in the previous section, finding optimal memory compiler input parameters in an efficient manner requires a model with low inference time to replace expensive compiler evaluations. Such a model accepts the same input parameters as the compiler it is trained to substitute during optimization. To refer to the set of model input parameters in contrast to compiler input parameters, we use the customary term ``explanatory variables''. From the set of explanatory variables, a behavioral model infers compiler \ac{PPA} outputs, so-called target variables.

Note that although memory compilers produce many essential artifacts such as netlists and RTL codes, only the task of predicting \ac{PPA} outputs is to be adopted by the behavioral model. Once chip designers have selected a compiler parametrization from optimizer results, they execute the actual memory compiler. Subsequently, they verify that the behavioral model's \ac{PPA} predictions are in line with compiler \ac{PPA} outputs and the remaining compiler artifacts are used for the product design flow.

A single behavioral model with sufficient capacity could in principle be fitted for all compilers together. However, routinely updated compiler versions as well as an expanding range of supported technology nodes would require frequent training of such a central model, a process which is computationally expensive. Therefore, we maintain a separate model for each compiler, creating what is commonly referred to as a ``model zoo'' \cite{Erickson.2017}. Models are added whenever new compilers or updated versions are released so that the model zoo contains one model per compiler and per compiler version. Because models are frozen after training, this approach also ensures that predictions for a given compiler version are consistent over time.

The behavioral models discussed here are trained through a supervised learning approach \cite{Friedman.2001}. In supervised learning, a model is fitted based on training data where each observation has known values for both explanatory variables ($x$) and target variables ($y$). Training data is obtained by executing the respective memory compiler for a given compiler parametrization. The memory compiler's \ac{PPA} outputs are referred to as ground truth data or $y$. In contrast, target variables inferred by the behavioral model are called predictions, or $\hat{y}$. Because the target variables are real-valued rather than discrete, the problem can be characterized as a regression task \cite{Friedman.2001}.

To collect ground truth data from the memory compiler, samples must first be obtained from the compiler parameter space. Exhaustive sampling of this vast combinatorial space is unfeasible, mainly owing to the compiler input parameters word width and word depth, which can each take on hundreds of possible values. To obtain the set of explanatory data (i.e. concrete values for explanatory variables), we randomly select 500 combinations of architectural and system parameters by sequentially fixing the value of each input parameter with uniform random probability among the valid choices.

Once compiler results are available, the respective behavioral model is trained and evaluated. Based on the model's prediction error on test set data, which is not used for training, we assess whether more training data is required. If that is the case, generation of another parametrization batch of 500 observations is triggered. Prior to sending the parametrization batch to the compiler, we remove those memories which are within size ranges with sufficiently low prediction errors based on separate prediction error analysis for different memory sizes (see Section \ref{sec:model-evaluation}). This results in decreasing parametrization batch sizes towards the end of the iterative training process, consequently speeding up the data generation process.

Because the amount of training data generated is driven by the quality of the model's predictions, the size of dataset differs between behavioral models. In our current model zoo for 25 compilers across 3 technology nodes, the median total number of observations is 2,500, with some models needing as little as 500 observations and others requiring up to 6,000 observations to reach satisfactory prediction errors. For training data generation, the memory compilers are executed in parallel threads. Different compilers have different run times and parallelization resources; typically, 20 compilers are run in parallel for 30 to 60 minutes per parametrization. The generation of a single parametrization batch of 500 observations therefore takes approximately 12 to 24 hours.

The nature of the collected data is summarized in Table \ref{tab:input-variables} and Table \ref{tab:target-variables} by example of a prototypical compiler. For all observed compilers, the explanatory variables (or compiler inputs) are discrete, ordinal values. With the exceptions of word depth and word width, these two to ten variables can take on between two and four possible values.

On the other hand, \ac{PPA} values are real-valued. We observe up to 20 different design corners. In total, the number of \ac{PPA} variables is given by the formula $c \times 5 + 1$, where $c$ is the number of design corners, 5 is the number of variables which are measured per design corner (read power, write power, leakage, cycle time, and access time), and 1 represents the variable ``area'', which is measured independently of design corners. For the average case of 15 design corners, the formula indicates 76 target variables.

\begin{table}[]
  \centering
  \caption{Input parameters of a prototypical memory compiler}
  \label{tab:input-variables}
  \begin{tabular}{lllll}
  \textbf{\begin{tabular}[c]{@{}l@{}}Parameter\end{tabular}} & \textbf{\# Choices} & \textbf{Range / values} & \textbf{} & \textbf{} \\
  Bank                                                                & $4$                        & $[1, 8]$                &           &           \\
  Column mux                                                          & $3$                        & $[4, 16]$               &           &           \\
  Periphery \acs{VT}                                         & $3$                        & $\{\textrm{low} < \textrm{standard} < \textrm{high}\}$ &           &          \\
  Redundancy                                                          & $3$                        & $\{\textrm{None} < \textrm{Row} < \textrm{Row + IO}\}$ &           &           \\
  Word depth                                                          & $223$                      & $[32, 32768]$           &           &           \\
  Word width                                                          & $313$                      & $[8, 320]$              &           &
  \end{tabular}
\end{table}

\begin{table}[]
  \centering
  \caption{\ac{PPA} dimensions of a prototypical memory compiler}
  \label{tab:target-variables}
  \begin{tabular}{lllll}
  \textbf{\begin{tabular}[c]{@{}l@{}}Parameter\end{tabular}} & \textbf{Range (approx.)} & \textbf{Unit}        & \textbf{} & \textbf{} \\
    Area                                                                & $(0, 10^5]$             & ${\mu}m^2$           &           &          \\
    Access time                                                         & $(0, 2]$                & $ns$                 &           &           \\
    Cycle time                                                          & $(0, 3]$                & $ns$                 &           &           \\
    Dynamic power (read)                                                & $(0, 30]$               & $\frac{{\mu}A}{MHz}$ &           &           \\
    Dynamic power (write)                                               & $(0, 30]$               & $\frac{{\mu}A}{MHz}$ &           &           \\
    Leakage                                                             & $(0, 10^5]$             & ${\mu}A$             &           &
  \end{tabular}
\end{table}

For this work, we model compiler \ac{PPA} outputs by means of fully connected feed-forward neural networks. Although the use of other regression models is conceivable, we propose feed-forward neural networks for several reasons. Firstly, their flexible structure allows their application to arbitrarily complex problems. Secondly, their ability to capture highly non-linear relationships is well-suited for modelling \ac{PPA} from compiler inputs. Lastly, neural networks allow analytical computation of gradients of the target variables with respect to explanatory variables, a property we deem highly desirable for future work on yet more challenging optimization tasks (see Section \ref{sec:outlook}).

We proceed to discuss data preparation, which is a significant factor contributing to the successful application of any machine learning technique \citep{Kotsiantis.2006}. As input to neural networks non-numerical input variables must be encoded. All explanatory variables are ordinal, and as such can be encoded as integer values. We further add the hand-crafted explanatory variable ``size'' to the set of explanatory variables, which is the product of word width and word depth following the assumption that it is a better predictor of the target dimensions area, leakage, and dynamic power than either of word depth or word width alone. Target variables, on the other hand, are transformed by applying the square root. This transformation aims to give more weight to observations with smaller targets, where deviations of the same absolute magnitude have a relatively larger effect. Lastly, to align variances, explanatory variables as well as target variables are standardized to a mean of zero and a standard deviation of one (also referred to as z-score normalization). Subsequently, min-max normalization is applied to ensure that explanatory and target variables of all observations are within a range of $[-1, 1]$ (see also \citep{Kotsiantis.2006}). The mean, standard deviation and extrema used for standardization and normalization are estimated based on the training set prior to training. Before inference, those same scaling factors are used for rescaling explanatory variables to the transformed scale and for transforming predictions back to the original scale thereafter.

Neural network-based models are highly sensitive to the selection of a suitable architecture and training procedure \citep{Domhan.2015}. The reader is referred to Section \ref{sec:model-evaluation} for experimental justification of the choices laid out in the following. On the network's input layer, the number of units is equal to the number of compiler input parameters of the respective memory compiler. On the ouput layer, one unit is present for each target variable. Between input and output layers, there are two hidden layers. The number of units on each hidden layer is set to be equal to eight times the number of input units, which allows the network's capacity to scale with the problem complexity as implied by the number of compiler input parameters. Input and hidden layers are activated using the sigmoid function, whereas no activation function is used for the output layer.

For training, the dataset is split randomly into approximately two thirds for training and one third hold-out data. The hold-out set is again split into two thirds for evaluation (i.e. the test set) and one third for validation. The validation set is used to determine when training should be stopped, a technique called ``early stopping'' \citep{Caruana.2001}. Early stopping is applied by checking validation set performance every 200 epochs, where one epoch encompasses one full pass over the training set during which network weights and biases are updated. If ten sequential checks find no reduction of prediction error (within a tolerance of $1 \times 10^{-3}$), training is stopped. Updates of network weights and biases are determined using the Adam optimizer \citep{Kingma.2014} with a learning rate of $1 \times 10^{-3}$ and the mean absolute error is used as the loss function. The update steps are computed based on so-called mini-batches \citep{Kingma.2014} of 100 observations, which are sampled from the training set with uniform random probability and without replacement. Note that these mini-batches are completely unrelated to the parametrization batches used for training data generation; instead, mini-batch is a standard term referring to sets of samples from the available training data used to compute a stochastic gradient during neural network training.

\section{Evaluation}
\label{sec:evaluation}
To qualify a model for real-world predictions of unseen data, its reliability must be assessed in the context of its application. This section discusses the acceptance criteria defined for behavioral models and the memory optimizer. We further verify our approach and model architecture by comparing prediction errors across different neural network architectures as well as other state-of-the-art machine learning algorithms.

\subsection{Acceptance Criteria}
A model's quality can unbiasedly be estimated by evaluating its prediction error for unseen data (held out during training). Section \ref{sec:model-evaluation} discusses the appropriate error metrics and presents exemplary results of model performance for one of the models trained for this study. The behavioral model chosen for the analyses presented in \ref{sec:model-evaluation} is representative of our model zoo in terms of number of explanatory and target variables; with 4500 observations, the size of dataset is above average.

Although the prediction error is an important aspect, it fails to capture the impact on actual decision-making based on the memory optimizer. Because the memory planning workflow includes a final collection of accurate PPA outputs from the memory compiler, inaccurate predictions are tolerable as long as they lead to the selection of the correct memory parametrization. Section \ref{sec:decision-quality} therefore discusses and estimates the reliability the memory optimizer in the context of decision making.

\subsection{Model Evaluation}
\label{sec:model-evaluation}
To summarize the quality of a model's predictions, we compare predictions for observations from the hold-out set to the respective ground truth values, i.e. compiler \ac{PPA} output. However, directly using the absolute prediction error has two disadvantages: On the one hand, interpretation is difficult because the error magnitude depends on the scale of the target variable. Comparing the quality of predictions between dimensions of different units, for example, is not possible. On the other hand, comparing the absolute prediction error between two observations may not be adequate. This is illustrated by two observations for which the absolute prediction error is the same, whereas the ground truth is different; the impact of a large prediction error (i.e. the nominator) is more significant when the true target (i.e. the denominator) is smaller. We therefore use a relative prediction error to scale the absolute difference by the magnitude of the true target value. A naive approach to this is the \ac{APE}, which is computed as shown in Equation \ref{eq:ape}.

\begin{equation}
  \label{eq:ape}
  \text{\ac{APE}} = \frac{\hat{y} - y}{y} \times 100
\end{equation}

While \ac{APE} is popular because of its numerous benefits, such as scale-independence, ease of interpretation, and scaling with the true target value, past work has also uncovered severe issues with the metric, one of which is the systematic bias towards models which under-predict \citep{Tofallis.2014}. To avoid these drawbacks, \cite{Morley.2018} introduced the symmetric signed percentage bias based on the log accuracy ratio (see also \cite{Tofallis.2014}). This metric does not favor over- or under-predictions, yet it is easily interpretable as a percentage. As we are interested in the magnitude of errors rather than the direction, we remove the sign by taking the absolute value. Henceforth we use the unsigned symmetric percentage bias as shown in Equation \ref{eq:spb}, referring to it as relative error or percentage error.

\begin{equation}
  \label{eq:spb}
  \text{Symmetric Percentage Bias} = \left( exp \left( \left| log\left( \frac{\hat{y}}{y} \right) \right| \right) -  1\right) \times 100
\end{equation}

Because the relative error is lower bounded by zero, but unbounded upwards, its distribution is typically skewed, exhibiting a tail on the right hand side. This feature makes the arithmetic mean, which is heavily influenced by outliers, inappropriate as a measure of average. To summarize the prediction error across observations, we therefore use the median to obtain a more representative measure of average. Note that usage of the arithmetic mean to average multiple medians is robust to the skew and more expressive, for example when aggregating across target variables.

\begin{figure}[!htb]
  \centering
  \begin{minipage}[t]{.48\textwidth}
    \centering
    \includegraphics[width=1\textwidth]{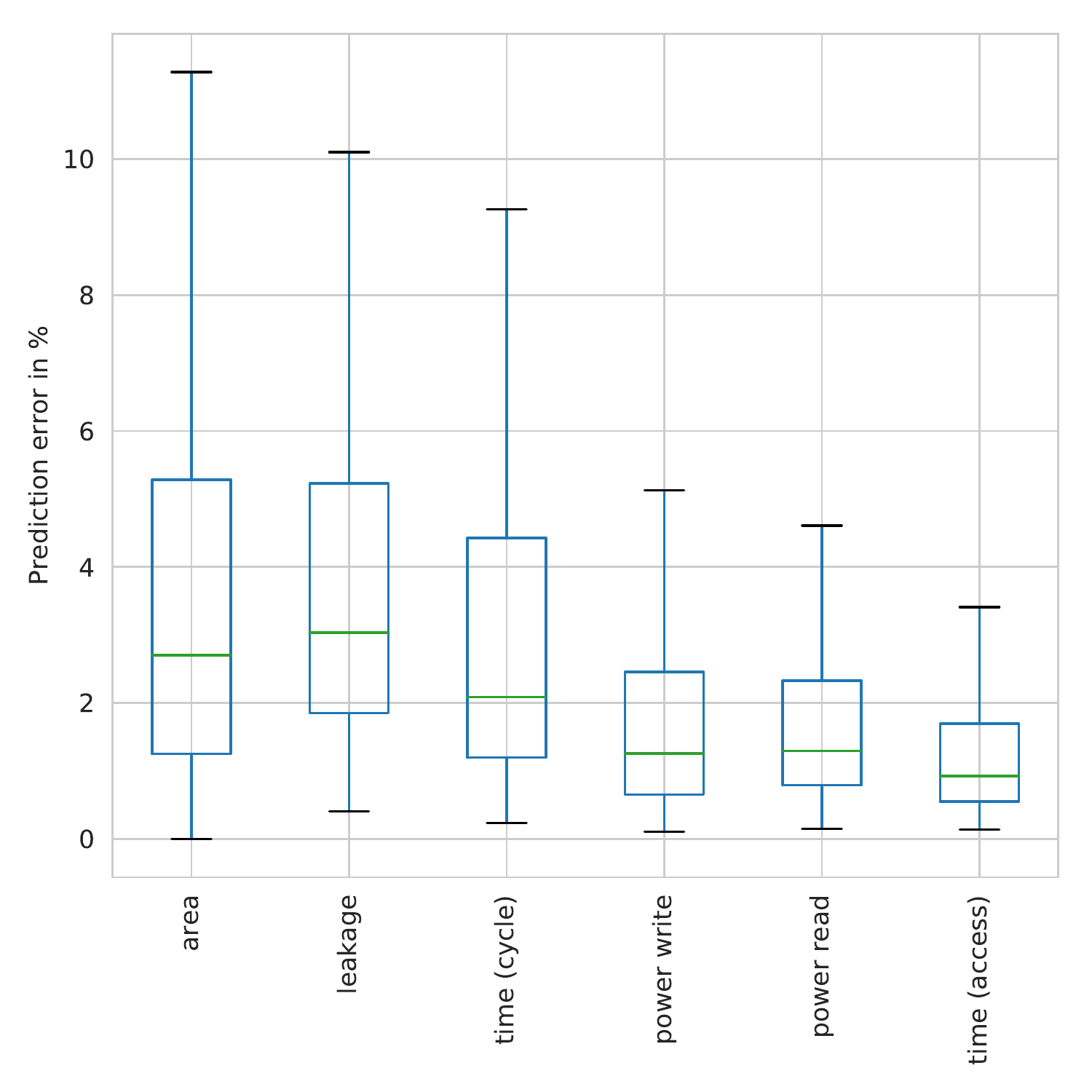}
    \caption{Relative error on the test set of one representative model, grouped by \ac{PPA} dimension. Box edges represent the 25\%- and 75\%-quantiles, the distance between them is the interquartile range. Lines in the center of each box represent the median. Whiskers on each side of a box extend to include all observations which are within a distance of 1.5 times the interquartile range from the 25\%- and 75\%-quantiles respectively. Outliers outside of this range are not shown.}
    \label{fig:avg-rel-loss-per-var}
  \end{minipage}%
  \hfill
  \begin{minipage}[t]{.48\textwidth}
    \centering
    \includegraphics[width=1\textwidth]{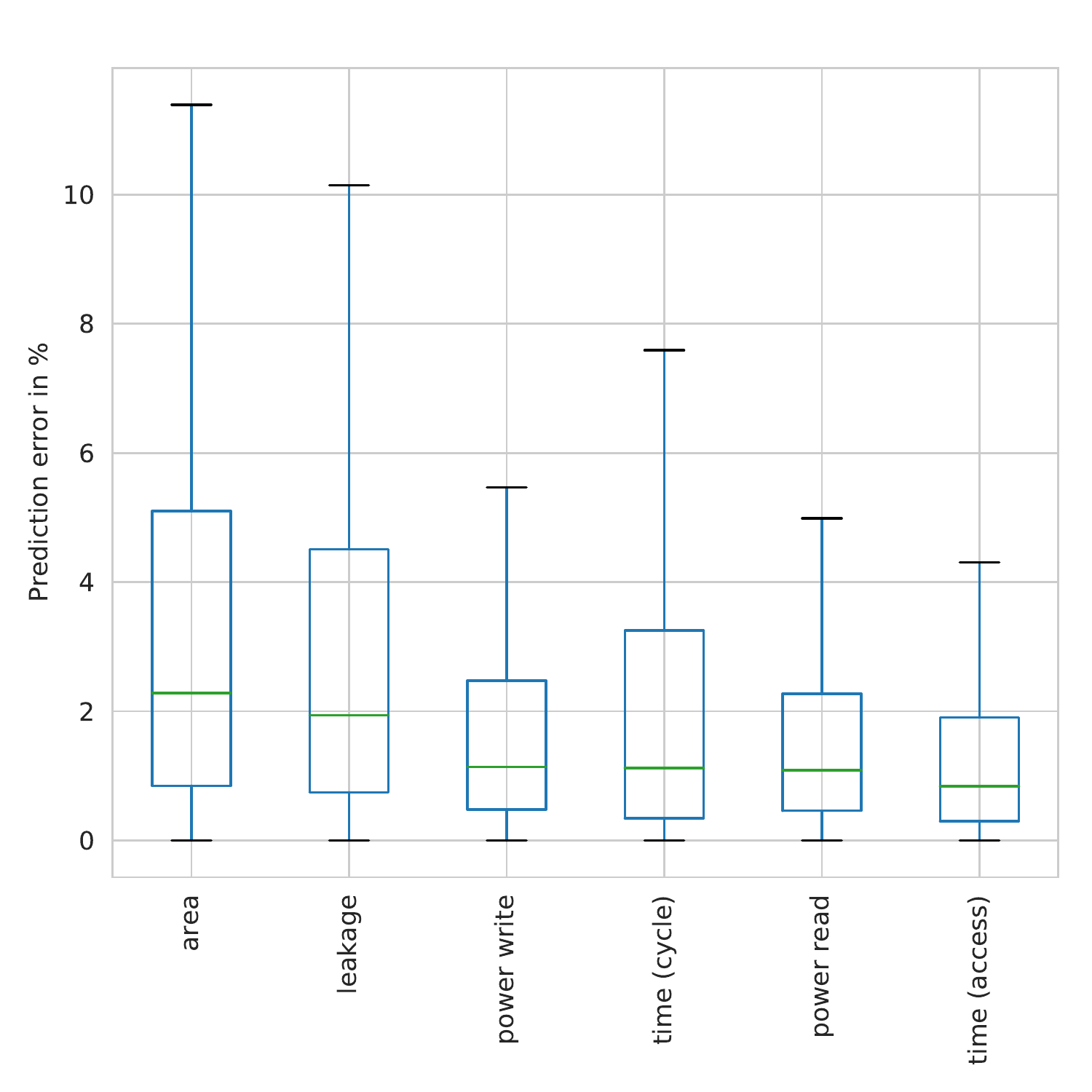}
    \caption{Relative test set error across all 25 models grouped by \ac{PPA} dimension, with equal weight attributed to each model.}
    \label{fig:rel-loss-across-compilers}
  \end{minipage}%
\end{figure}

To understand the reliability of each dimension's predictions, we break down the prediction error by target variable. As explained in Section \ref{sec:optimizer}, some \ac{PPA} dimensions (leakage, access time, cycle time, and dynamic read and write power) are represented by many \ac{PPA} variables for different design corners. We aggregate the model's target variables into \ac{PPA} dimensions in order to obtain a less cluttered visualization. In contrast to the aggregation across observations, the arithmetic mean is used to average across variables. Figure \ref{fig:avg-rel-loss-per-var} shows the result of this analysis for a single behavioral model. The same analysis is performed across all behavioral models, the result of which is shown in Figure \ref{fig:rel-loss-across-compilers}. In order to give each model equal influence on the displayed summary statistics, we randomly sample the same number of observations from each model's test set before computing and aggregating all errors.

\begin{figure}[!htb]
  \centering
    \includegraphics[width=.55\textwidth]{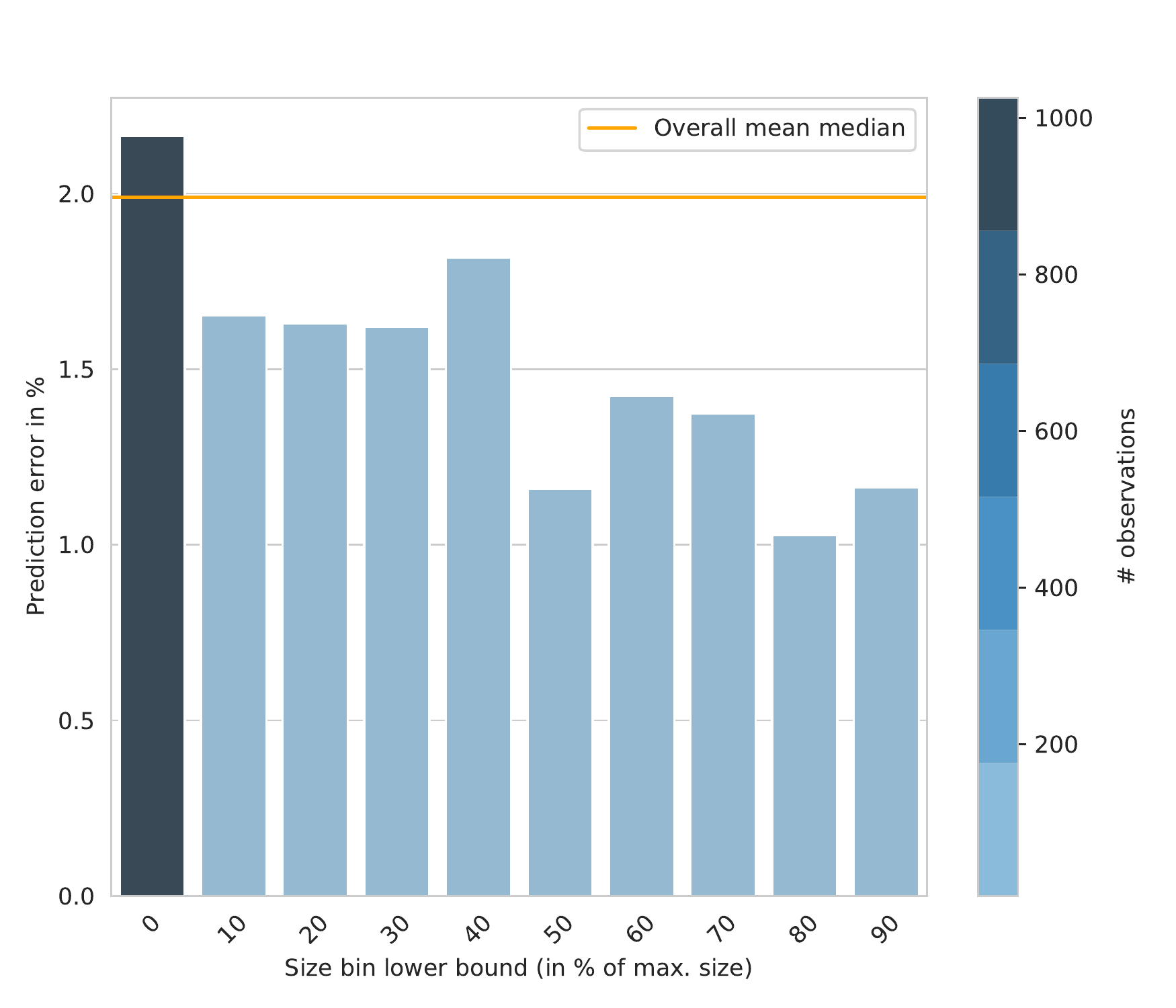}
    \caption{Average relative error evaluated on the test set averaged for different memory size ranges. Bars represent size ranges, and their shade indicates the number of test set observations available in that size range.}
    \label{fig:avg-rel-loss-per-size}
\end{figure}

The figure shows that median errors of at most 3\% are achieved across variables. Best case predictions for each variable exhibit close to no deviation from ground truth. The 25\%-quantile is consistently between 0\% and 2\% error, indicating a very low prediction error for a quarter of observations in the test set. The 75\%-quantile is below 3\% for 3 out of 6 \ac{PPA} dimensions, while it is just above 5\% for area and leakage, for which prediction error is highest.

We further analyze the relative prediction error in relation to memory size in bits. This is done to focus training data generation on size ranges with insufficient prediction quality as discussed in Section \ref{sec:behavioral-model}. For the analysis, observations from the test set are first grouped into 10 bins according to their size in bits. The average per bin is then computed by calculating the median across observations, before aggregating across variables using the arithmetic mean. Figure \ref{fig:avg-rel-loss-per-size} reveals that the expected prediction error is distributed fairly evenly across bins, asserting that prediction quality is stable for arbitrarily sized instances, with average prediction errors consistently below 2.5\%. 

As the shade of each bar indicates the number of observations in the respective bins, it is visible that most observations are small memories. When the number of bit cells is small, the impact of the memory periphery on a memory's \ac{PPA} is relatively larger. As most compiler input parameters affect the periphery rather than the bit cell array, it is more difficult to accurately predict the \ac{PPA} of small memories. To achieve a comparable prediction error, more training data is thus needed for small word widths and word depths.

 To determine an optimal architecture for the neural network, we perform a grid search across neural network architecture options. We test different values for the following options: the number of hidden layers, the hidden layer unit multiplier, the output layer activation function and the hidden layer activation function. For the number of hidden layers, we test the values $\{1, 2, 4, 6, 8\}$. The hidden unit multiplier, which determines the number of units on each hidden layer by means of multiplication with the input layer dimension, the values $\{1, 2, 4, 6, 8, 10\}$ are considered. We skip extremely large network architectures where the number of hidden layers is eight and the hidden layer unit multiplier exceeds six. For hidden layers, the activation functions sigmoid, tanh and rectified linear (relu) are considered. Regarding the output layer, no activation and relu are tested. For the latter, the output range is corrected to be lower bounded by -1 instead of 0 so that the same data scaling factors, which are based on values in $[-1, 1]$, can be used (see Section \ref{sec:behavioral-model} for details on data scaling). All training parameters were held constant, setting the initial learning rate of the Adam optimizer \citep{Kingma.2014} to $1 \times 10^{-3}$, the mini-batch size to 100, disabling dropout and applying early stopping as described in Section \ref{sec:behavioral-model}. In total, we evaluate 180 different neural network architectures.

\begin{figure}[!htb]
  \centering
  \includegraphics[width=1\textwidth]{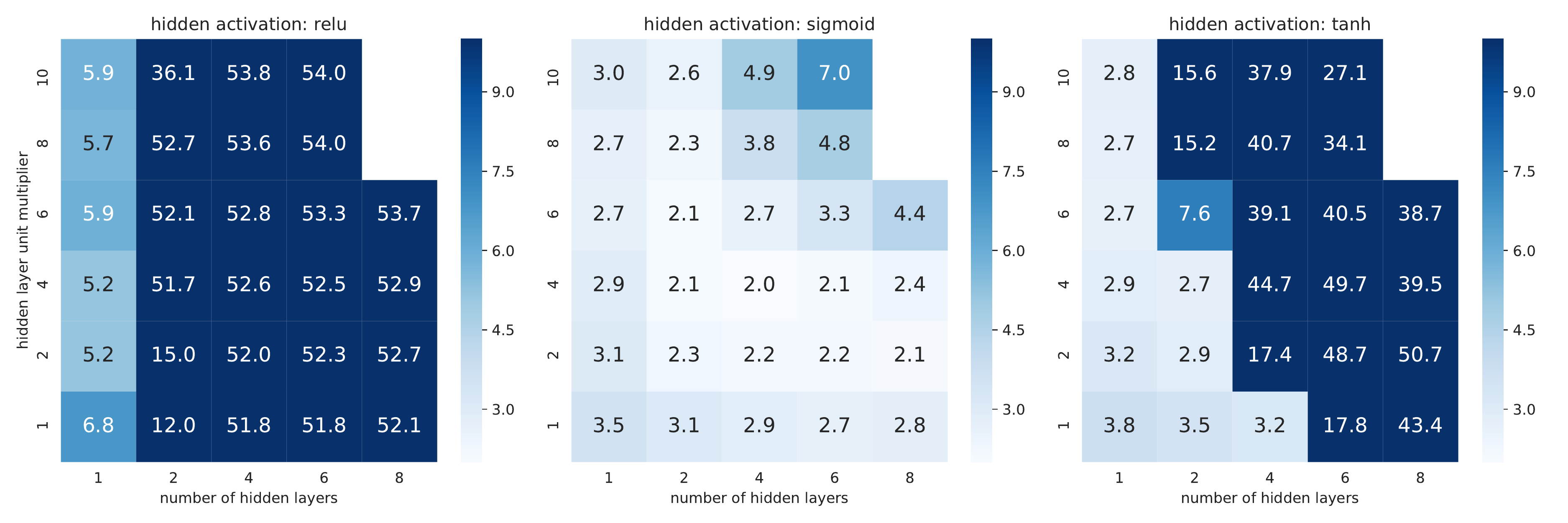}
  \caption{Average, cross-validated relative prediction errors for different neural network architectures. Architectures evaluated in each plot share a hidden layer activation function, as labeled above the respective chart. Squares within each plot correspond to a specific number of hidden layers (x-axis) and a hidden layer unit multiplier (y-axis), which determines the number of units on each hidden layer through multiplication with the number of input units.}
  \label{fig:grid-search-architecture}
\end{figure}

 Figure \ref{fig:grid-search-architecture} shows one plot for each hidden layer activation function. Within each plot, there is one square for each evaluated architecture, labeled with the relative prediction error. This prediction error is computed by taking the median relative error per variable, before using the arithmetic mean to aggregate across variables. Additionally, this metric is averaged across the folds of three-fold cross-validation using the arithmetic mean. In three-fold cross-validation, the dataset is split three times, each time using a different portion of data for training and the remainder for testing. The color scale of the charts ends at a prediction error of 10\% in order to facilitate discernibility among lower values. Results obtained using relu as the output activation are omitted because they consistently exhibit prediction errors of more than 30\%, with a single exception (2.9\% prediction error at eight hidden layers and a hidden unit multiplier of one). All plots shown therefore use no activation function for the output layer.

The most obvious trend visible in Figure \ref{fig:grid-search-architecture} is that the sigmoid activation of hidden layers tends to yield the best results. The tanh activation achieves appreciable results, which, however, cannot be sustained for larger network architectures. The relu activation for hidden layers, on the other hand, appears inadequate, with prediction errors consistently above 5\%. The lowest prediction error found overall is 2\%, which is achieved with a network architecture of four hidden layers, sigmoid activation for hidden layers, and four times as many hidden units as input units. Similar scores are obtained for many other configurations which use sigmoidal hidden layer activations, especially when at least two hidden layers are present and the hidden layer unit multiplier is at least two. The trend appears to decline when both of these neural network architecture parameters approach the upper end of the tested range. Based on preliminary results of this analysis, a neural network architecture with two hidden layers and a hidden layer unit multiplier of eight (with an expected prediction error of 2.3\%) has been used for other analyses in this work.

Aiming to understand which correlations the behavioral model has learnt from the data, we analyze the derivative of the neural network with respect to the explanatory variables. A positive derivative indicates a positive correlation between explanatory variable and target variable, and vice versa. The derivatives are computed and averaged across the combined training and test set using the arithmetic mean, which is also used to aggregate target variables to \ac{PPA} dimensions. For each explanatory variable, the gradient values are normalized into a range of $[-1,1]$ in order to emphasize which \ac{PPA} dimensions are most impacted by each explanatory variable.

\begin{figure}[!htb]
  \centering
  \includegraphics[width=0.7\textwidth]{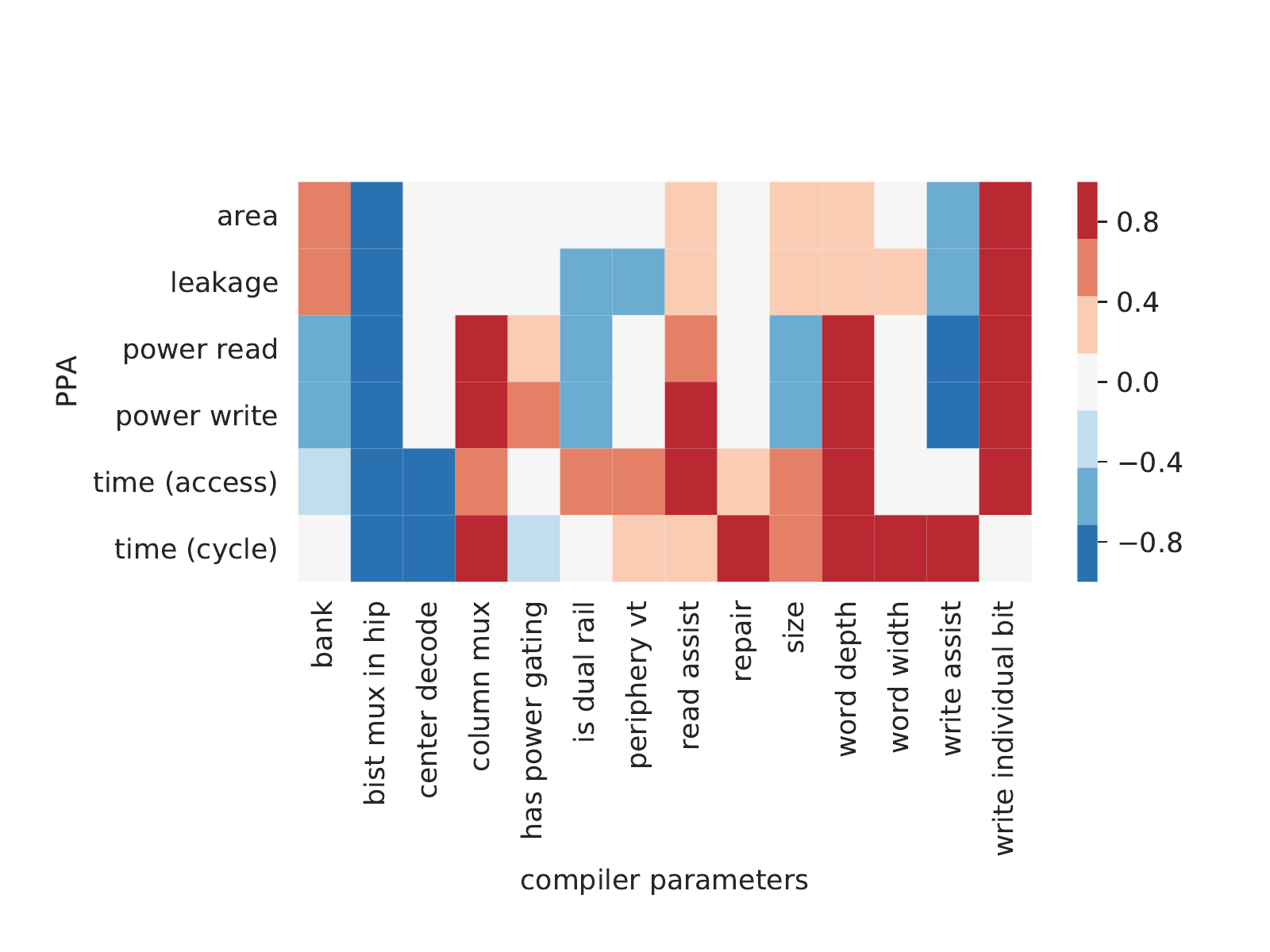}
  \caption{Average gradient of neural network outputs (\ac{PPA} variables) with respect to compiler input parameters. The values are normalized by input variable, so that each column shows which \ac{PPA} variables are impacted most by the respective compiler option.}
  \label{fig:feature-importance}
\end{figure}

The results of this analysis, visualized in Figure \ref{fig:feature-importance}, show the predicted impact of compiler input parameters on \ac{PPA}. For example, the model seems to have properly learnt that as the periphery voltage threshold increases, leakage decreases, but cycle time increases. A larger number of banks, on the other hand, leads to a larger area while decreasing access time. Another relationship learnt by the model is unsurprising: there is a positive correlation between size in bits and area.

We further compare our approach of fitting compiler data using feed-forward neural networks to state-of-the-art regression techniques from statistics and machine learning. Specifically, we compare least squares linear regression, gradient boosting \citep{Friedman.2001-1}, AdaBoost \citep{Freund.1997}, random forest regression \citep{Breiman.2001}, and polynomial regression (linear regression with polynomially transformed explanatory variables). We evaluate the prediction error of different configurations of these models using cross-validation. Prior to applying cross-validation, we extract a portion of the data as a validation set used for early stopping neural network training. For the neural network-based model, we use the same neural network architecture as for other analyses is this work.

\begin{figure}[!htb]
  \centering
  \includegraphics[width=0.6\textwidth]{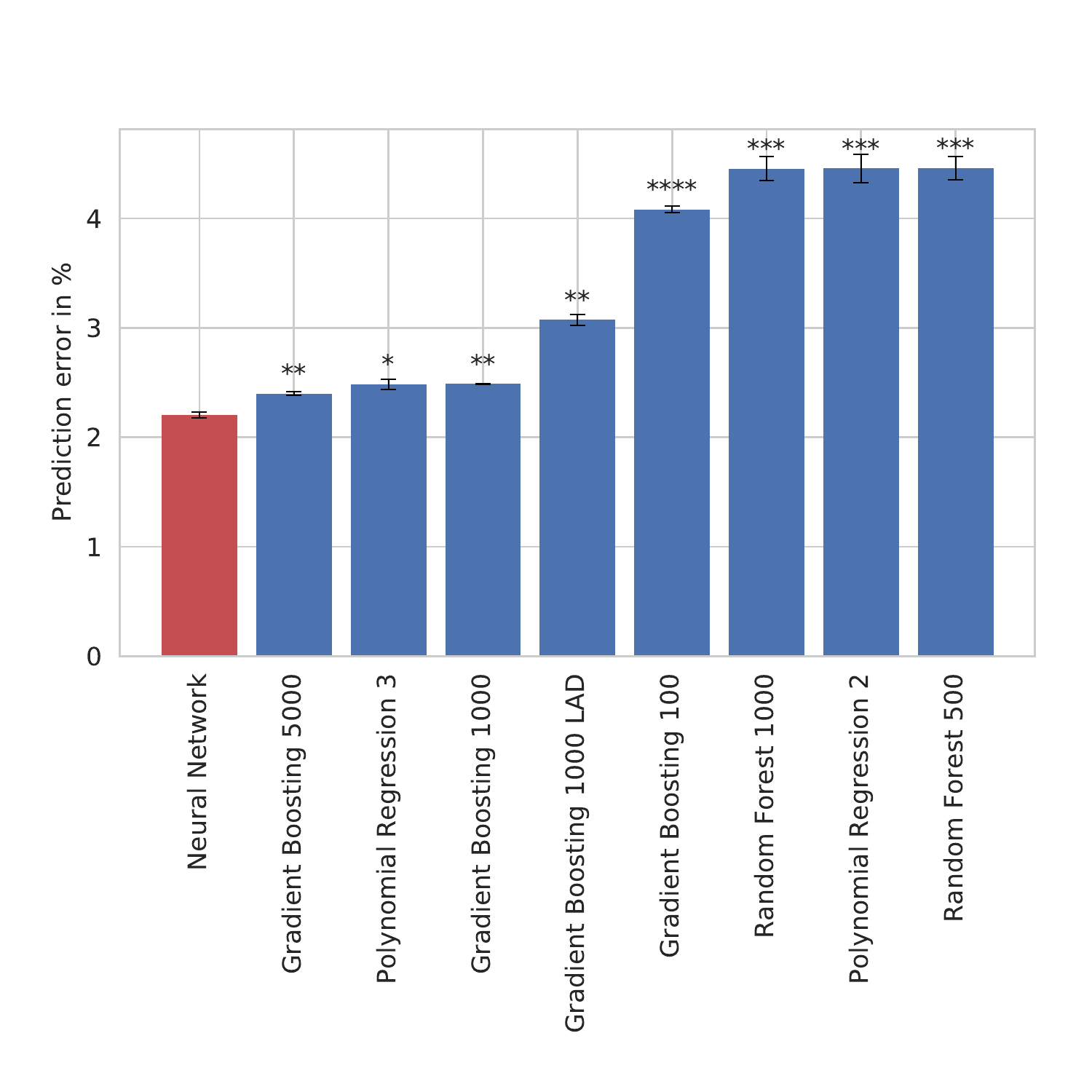}
  \caption{Comparison of relative prediction errors between various regression techniques and the proposed feed-forward neural networks. Each bar represents the mean across three cross-validated train-test cycles. Whiskers indicate the standard deviation. All methods are regression models, not the sometimes better known classification techniques with the same name.}
  \label{fig:regression-comparison}
\end{figure}

The results are presented in Figure \ref{fig:regression-comparison}, where linear regression as well as AdaBoost with 500 and 1000 estimators are excluded for clarity as their errors exceed 6\%. The figure clearly shows that our proposed regression using neural networks is the best approach among the compared techniques. Seven out of twelve approaches (including those three not shown in the chart) are outperformed by more than one percent, while closer followers like gradient boosting with 500 estimators and 3rd degree polynomial regression are outperformed by a small, yet significant margin. Statistical significance was determined using a two-sided, paired Student's t-test \cite{Student.1908} to test the null hypothesis that the true mean error of each method is the same as the true mean error of the proposed method. The null hypothesis is rejected for every evaluated method, where asterisks above each bar represent the significance level. The number behind a method's name indicates the number of estimators (regression trees) in case of ensemble methods, or the polynomial degree in case of polynomial regression.  Gradient boosting regression was performed using least squares loss or least absolute deviation; use of the latter is indicated by ``LAD'' in the figure. For unspecified parameters, the default choices as provided by the scikit-learn \cite{Pedregosa.2011} framework at version 0.20.3 are used.

The low inference time of neural networks is among the key arguments of their application to the memory selection problem. Table \ref{tab:network-timing} shows the inference time of a prototypical model from our model zoo, measured for different numbers of samples predicted. This is the pure prediction time, excluding the overhead of several set-up tasks, such as loading the model from disk into memory and preprocessing the data. Timing analysis was performed on a machine with a single virtual CPU and 2 Gigabytes of memory. Inference was repeated 1,000 times, retaining the minimum run time in order to approximate the lower bound of required computing time. As the timing results demonstrate, inference takes consistently less than a second for up to 10,000 observations. Moreover, even inference of 100,000 observations does not take significantly longer than a second. The scaling factor shows how much longer the inference is for each sample size compared to the time taken to predict a single observation. This illustrates that predicting large numbers of samples at once is much more efficient than individual prediction, a property which enables the efficient search space evaluation upon which the optimization framework relies.

Training, which is performed on the same hardware as inference, approximately takes between 5 and 30 minutes. However, training time depends not only on the number of observations, explanatory variables, and target variables in the dataset, but also on the number of epochs needed until early stopping determines convergence. For example, training with a typical dataset of 2,500 observations for 500 epochs takes 40 seconds. Training is normally stopped after 5,000 to 20,000 epochs. Trained models are stored on disk and use less than 200 Kilobytes, but can easily be compressed to less than half that size.

\begin{table}[!htb]
  \centering
  \caption{Inference times of a prototypical behavioral model}
  \label{tab:network-timing}
  \begin{tabular}{lllll}
  \textbf{\# Samples}  & \textbf{Time in seconds}          & \textbf{Scale factor} \\
  $1$                         & $1.87 \times 10^{-4}$             & $1.00$  \\
  $10$                        & $2.27 \times 10^{-4}$             & $1.22$  \\
  $10^{2}$                    & $5.92 \times 10^{-4}$             & $3.17$  \\
  $10^{3}$                    & $1.89 \times 10^{-3}$             & $1.01 \times 10^{1}$  \\
  $10^{4}$                    & $8.12 \times 10^{-2}$             & $4.35 \times 10^{2}$  \\
  $10^{5}$                    & $1.11$                            & $5.91 \times 10^{3}$  \\
\end{tabular}
\end{table}

\subsection{Optimizer Evaluation}
\label{sec:decision-quality}
Prediction error evaluation alone does not provide a comprehensive view of the quality of optimizer results. The goal of this analysis is to devise a metric capable of estimating decision reliability in the context of the memory optimizer. For this purpose, assume that a chip designer will select the memory parametrization ranked first according to their selected \ac{PPA} criterion (see Section \ref{sec:optimizer}). Under this assumption, it is intuitive that the correct decision will be made if the truly best suited memory is ranked first.

\begin{figure}[!htb]
  \centering
  \includegraphics[width=0.4\textwidth]{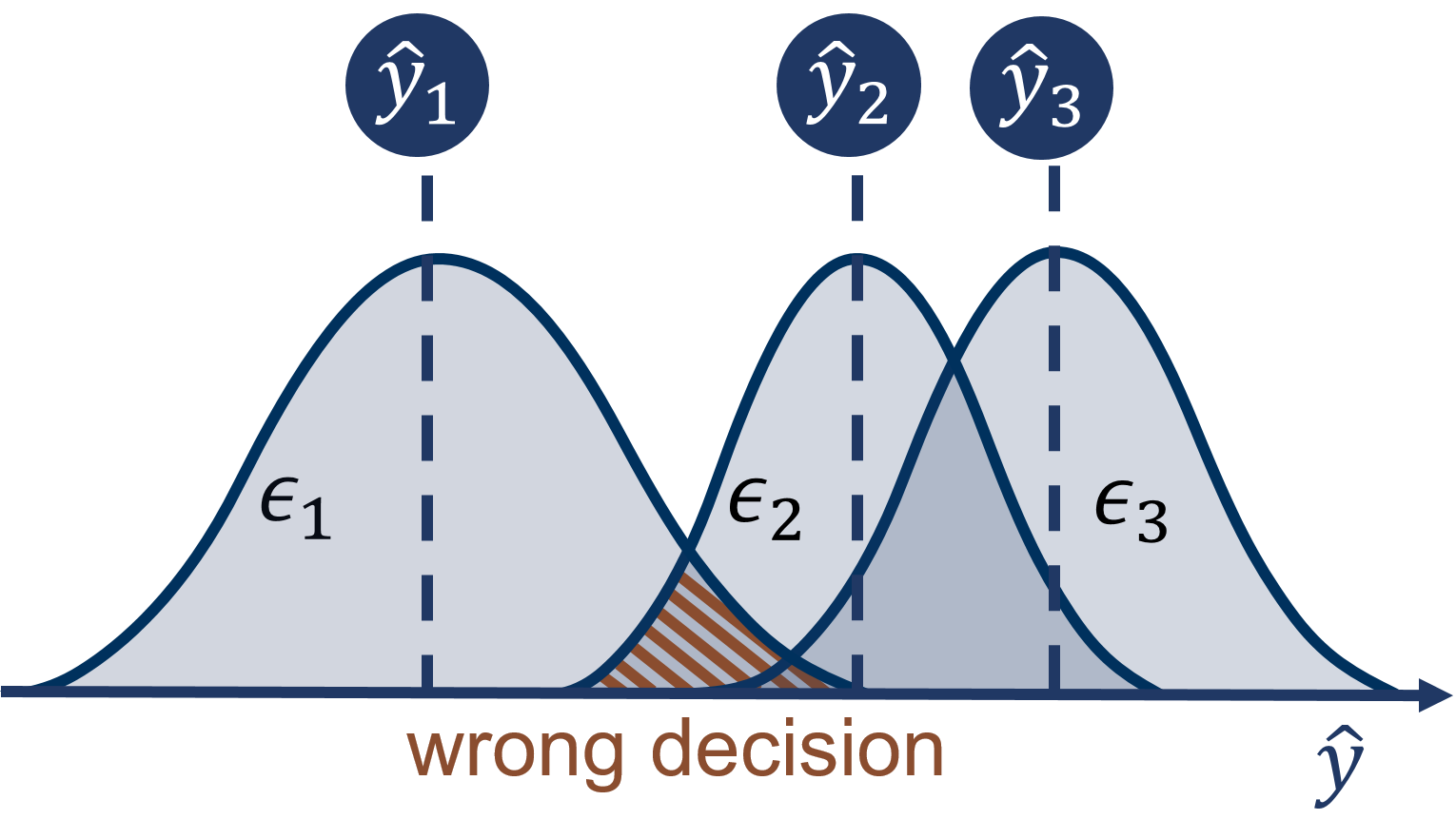}
  \caption{Illustration of the computational basis for the decision reliability analysis. When the error distributions $\epsilon$ centered around ranked predictions $\hat{y}$ overlap, the reliability of the ranking and hence of the ranking-based decision are impaired. The decision reliability score of a given ranking is computed by averaging the estimated size of the wrong decision regions of every pair of results involving the first ranked prediction.}
  \label{fig:decision-reliability}
\end{figure}

Assume that the compiler parametrizations ranked first ($x_1$) and second ($x_2$) in the optimizer results are separated by a distance of $d = \hat{y}_1 - \hat{y}_2$ in terms of predicted \ac{PPA}. As illustrated in Figure \ref{fig:decision-reliability}, it follows then that the two instances will be ranked in the wrong order (relative to each other) if the sum of the prediction errors $\epsilon_1 + \epsilon_2$ of the results exceeds $d$. In other words, when the over-estimation of the first ranked result and the under-estimation of the second ranked result exceed the distance between the two, a wrong decision will be made. The same applies to all pairs $(x_1, x_i)$ in a result's ranking.

Estimating decision reliability in the context of the memory optimizer poses some challenges. Firstly, evaluations can no longer be made on a per-compiler basis, as multiple memory compilers make up the set of results of a single optimizer run. Consequently, all compiler models which apply to the selected port configuration have to be jointly evaluated. Secondly, the vast number of possible parametrizations for any given optimizer run makes it infeasible to gather exhaustive ground truth \ac{PPA} outputs from the memory compilers. This problem becomes even more apparent when the goal is not to assess an optimization run for a single memory, but a large enough sample of such.

To approach a more feasible evaluation method, the expected prediction error - as estimated based on test set data - is devised to approximate each result's deviation from ground truth. We aim to estimate the average integral overlap of errors around the predicted results $\epsilon + x$, corresponding to the shaded area of Figure \ref{fig:decision-reliability}. As some errors may not be distributed normally, we adopt a numerical method of computing the integral by randomly sampling from the joint error distributions of a result set. This means that we modify the ranked results given by the optimizer by repeatedly sampling from each parametrization's expected error distribution and adding the sampled errors to the prediction. When predictions for all memories have been adjusted by the sampled errors, a new ranking is computed. We then determine if the initially first ranked instance is still ranked first. After repeating this process 1,000 times, we compute the proportion of repetitions where the first ranked instance remained the same. We interpret this proportion as a measure of decision reliability in the given optimizer run, where 100\% is the best attainable value and 0\% is the lower bound.

Expected error distribution of each result is estimated based on the prediction errors of at least 100 similar samples from the test set. Similar samples are chosen based on proximity in terms of size in bits. When available, memories of the same size are used for error estimation, otherwise we select 50 neighbors with a larger size and 50 neighbors with a smaller size. The Shapiro–Wilk test \citep{Shapiro.1965} is conducted to test the null hypothesis of normality of errors. For distributions where the null hypothesis is rejected (based on a p-value of $0.05$), we obtain Gaussian kernel density estimates \citep{Scott.1992} instead.

We repeat the analysis for 3,000 different memory sizes sampled randomly with uniform probability from the set of all possible word depth and word width combinations. The analysis is conducted for a port configuration which contains 9 different memory compilers.

\begin{table}[!htb]
  \centering
  \caption{Decision reliability score across 3,000 optimizer runs}
  \label{tab:optimizer-eval}
  \begin{tabular}{lllll}
  \textbf{\ac{PPA} Dimension} & \textbf{Mean}          & \textbf{95\%-Quantile} & \textbf{Minimum}  \\
  area              & $100\%$                 & $100\%$                     & $100\%$     \\
  leakage           & $99.9\%$                & $100\%$                     & $16.8\%$    \\
  power (read)      & $99.2\%$                & $100\%$                     & $47.8\%$    \\
  power (write)     & $99.4\%$                & $100\%$                     & $41.4\%$    \\
\end{tabular}
\end{table}

As can be seen in Table \ref{tab:optimizer-eval}, decision reliability as indicated by this analysis is very high throughout \ac{PPA} dimensions. The mean score of each dimension is very close to one. For 95\% of optimizer runs we observe a decision reliability score of over 99\% for every target dimension, making wrong decisions highly unlikely for optimizations based on the evaluated port configuration. Minimum scores are above 15\% for all variables. The \ac{PPA} dimension area has a minimum score of 100\% indicating extremely reliable decisions.

\subsection{Comparison to Expert Design}
In order to assess the real-world benefit of the proposed memory optimization framework, we compare an existing memory selection performed by human experts to a selection made based on the optimizer. The existing expert-based selection consists of 5,623 memories of various bit sizes, target frequencies and other fixed system parameters. We apply the optimizer to minimize area, while for each memory the same target frequency which constrained the experts' selection must be met. For each memory, we select the physical instance ranked first according to area. Subsequently, we compare both methods in terms of \ac{PPA} dimensions, based on the same design corners as the expert-based selection criterion. \ac{PPA} of each selection method is assessed by using compiler outputs where available or network predictions otherwise, summing the individual memories' \ac{PPA} to calculate \ac{PPA} for the whole selection. The differences in \ac{PPA} are reported as a percentage of the \ac{PPA} of the expert-based selection. Less than four hours are required to complete the optimization of all memories, while optimization by human experts took approximately 10 work weeks.

\begin{table}[!htb]
  \centering
  \caption{\acs{PPA} difference of optimizer-based vs. expert-based memory selection, relative to the expert-based selection.}
  \label{tab:expert-vs-optimizer}
  \begin{tabular}{lllll}
  \textbf{\ac{PPA} Dimension} & \textbf{Difference}\\
  area                        & $-14\%$\\
  leakage                     & $-13\%$\\
  dynamic power (read)        & $-10\%$\\
\end{tabular}
\end{table}

The analysis shows that all three optimization targets could be significantly diminished by at least 10\% by applying the proposed solution. This was achieved partly at the cost of decreasing the frequency of the memories, reducing the margin toward the target frequency which is satisfied nevertheless. The success can also be attributed to the exhaustion of the extremely large search space, which is unfeasible within time and resource limitations of human experts. This is especially true when experts have to rely on memory compilers, which take significant time to produce \ac{PPA} outputs. On the other hand, exhaustive exploration by the proposed approach takes less than four hours, which is a substantial decrease from the weeks required for manual selection.

\section{Conclusion}
\label{sec:conclusion}
In this section, we discuss the results presented in Section \ref{sec:evaluation} and their implications for \ac{IC} design. We further explore future challenges and research opportunities related to memory compiler \ac{PPA} optimization.

\subsection{Results Assessment}
The evaluation of the optimizer shown in Section \ref{sec:decision-quality} illustrates the effectiveness of our proposed solution at finding the best possible compiler input parameters given design requirements with a decision reliability of over 99\% on average. This achievement is owed to models with high prediction quality as revealed by average prediction errors below 2.5\% in light of a complex, high-dimensional relationship between compiler input parameters and \ac{PPA} outputs. 

Meanwhile, the optimization is a fully automatic process with remarkably low run times averaging at less than ten seconds. New compilers versions are further supported in a matter of days after their release, with quality assurance provided by the automation of data generation, model training and model evaluation cycles.

As comparison with expert-based memory selection for a real-world chip demonstrates, the proposed solution also attains sizable gains over careful manual optimization, yielding more than 10\% reduction in terms of area, leakage and dynamic power. It is important to note that the memory optimizer is already in full productive use for multiple large volume design projects and not merely a proof of concept. Successful completion of real-world, commercial chip design projects which relied on the memory optimizer further manifests the value add of our approach. Through the use of the memory optimizer, the complexity of the design process was reduced, resulting in an estimated 20\% of time-savings of the selection and parametrization process for \ac{IC} products.

\subsection{Outlook}
\label{sec:outlook}
While the difficult task of optimizing compiler input parameters for a given memory has been solved to our satisfaction, many challenges in the space of memory compiler parameter selection remain. One major topic is the optimal tuning of many memories in an ensemble, which could enable valuable use cases such as system optimization or optimization of compounds of multiple physical macros. For either case, optimizing the system of memories as a whole rather than individually enables unique solutions with \ac{PPA} trade offs between memories rather than individually balanced instances, leading to improved overall \ac{PPA}. The ensemble problem is characterized by a combination of possibly thousands of memories, resulting in a much larger search space. Even given sub-millisecond model inference times, minimization techniques beyond exhaustive search are required. We envision that neural network properties, such as analytical gradient computation, can be exploited to efficiently guide the search space exploration. Applied to compound memories, such a solution would fill the optimization gap for memory design edge cases such as extremely high-frequency or large-size macros. On the other hand, rapid optima search across all memories of an entire \ac{IC} would revolutionize the design flow as a whole, making early product \ac{PPA} estimation as well as final memory optimization automated, fast and accurate.

\bibliographystyle{apalike}
\bibliography{literature}

\end{document}